\documentclass[preprint,prc,aps,showpacs,amsmath,nofootinbib]{revtex4}

\begin{document}
\title{Spin observables and the determination of the parity \\ 
of $\Theta^+$ in photoproduction reactions.}
\author{K. Nakayama and W. G. Love}
\affiliation{Department of Physics and Astronomy, University of Georgia, 
Athens, GA 30602, USA }

\begin{abstract}
Spin observables in the photoproduction of the $\Theta^+$ are explored for
the purpose of determining the parity of the $\Theta^+$. Based on reflection
symmetry in the scattering plane, we show that certain spin observables in
the photoproduction of the $\Theta^+$ can be related directly to its parity.
We also show that measurements of both the target nucleon asymmetry and the $%
\Theta^+$ polarization may be useful in determining the parity of $\Theta^+$
in a model-independent way. Furthermore, we show that no combination of spin
observables involving only the polarization of the photon and/or nucleon in
the initial state can determine the parity of $\Theta^+$ unambiguously.
\end{abstract}

\maketitle


\vskip 0.5cm

%
\newpage 
The pentaquark $\Theta ^{+}$ was predicted by Diakonov, Petrov and Polyakov~%
\cite{Diakonov} in 1997 in the chiral soliton model as the lowest member of
an anti-decuplet of baryons. The recent discovery of this truly exotic
baryon ~\cite{LEPS,DIANA,CLAS,SAPHIR,neutrino,Airapetian,Aleev} has
triggered an intensive investigation aimed at a determination of its basic
properties. Currently, available data do not allow for the determination of
either its spin or its parity. Moreover, theoretical predictions of these
quantum numbers, and especially the parity, are largely controversial. For
example, the quenched lattice QCD calculations \cite{lattice} identified the
spin 1/2 $\Theta ^{+}$ as an isoscalar negative parity state (see, however,
a recent quenched lattice QCD calculation with exact chiral symmetry \cite%
{Chiu}, where a positive parity is predicted for $\Theta ^{+}$). Also, QCD
sum-rule calculations~\cite{QCDsum} predict a spin 1/2 negative parity
state. In contrast, chiral/Skyrme soliton models ~\cite{Diakonov,Weigel} and
many other models~\cite{QUARK} predict a spin 1/2 positive parity isoscalar
state. There also exist theoretical studies which explore the possibility of
determining the quantum numbers and especially the parity of $\Theta
^{+}(1540)$ experimentally \cite{MODELS,Zhao,Nak,Oh}. In particular, spin
observables such as photon asymmetry and spin-correlation functions are
shown to be very sensitive to the parity of $\Theta ^{+}$ \cite{Zhao,Nak,Oh}%
. However, all these analyses rely on the particular model(s) used. A number
of authors have also carried out model-independent analyses aimed at an
unambiguous determination of the $\Theta ^{+}$ parity in both hadronic \cite%
{M-IND-H} and electromagnetic \cite{Rekalo} induced reactions.

In the present work, we perform a model-independent analysis of the $\gamma
N\rightarrow \bar{K}\Theta ^{+}$ reaction and show that certain spin
observables can be related directly to the parity of $\Theta ^{+}$. We also
show that measurements of both the target nucleon asymmetry and the $\Theta
^{+}$ polarization may be useful in determining the parity of $\Theta ^{+}$
unambiguously. Furthermore, we show that no combination of spin observables
involving only the polarization of the photon and/or nucleon in the initial
state can pin down the parity of $\Theta ^{+}$ in a completely
model-independent way. To obtain these results, we first derive the most
general spin structure of the reaction amplitude for both the positive and
negative parity $\Theta ^{+}$. Here we extract the spin structure of the
reaction amplitude following the method used in Ref.\cite{NL}, which is
based on its partial-wave expansion. The method is quite general and, in
principle, can be applied to any reaction process in a systematic way.
Usually, the structure of a reaction amplitude is derived based solely on
symmetry principles; the advantage of the present method is that it yields
the coefficients multiplying each spin operator in terms of the partial-wave
matrix elements. Details of the derivation will be reported elsewhere. In
what follows, we consider the $\Theta ^{+}$ to be a spin-1/2 baryon.
Hereafter, the superscript $\pm $ on any quantity (other than $\Theta $)
stands for the positive $(+)$ or negative $(-)$ parity of $\Theta ^{+}$.

For a positive parity $\Theta ^{+}$, the reaction amplitude takes the form 
\footnote{%
Actually, there is an issue of the parity (more precisely the relative
parity) of the kaon not being known. For a recent discussion see Ref.\cite%
{Pak}. Throughout this work, we assume the kaon to be a pseudoscalar meson.
If the parity of the kaon happens to be positive, all the results in this work
referred to be for positive parity $\Theta ^{+}$ should be interchanged with
those for negative parity $\Theta ^{+}$.} 
\begin{equation}
\hat{M}^{+}=F_{1}\ \vec{\sigma}\cdot \vec{\epsilon}+iF_{2}\ \vec{\epsilon}%
\cdot \vec{n}+F_{3}\ \vec{\sigma}\cdot \hat{k}\vec{\epsilon}\cdot \hat{q}%
+F_{4}\ \vec{\sigma}\cdot \hat{q}\vec{\epsilon}\cdot \hat{q}\ ,
\label{SSTRUC_P}
\end{equation}%
where $\hat{k}$ and $\hat{q}$ are unit vectors in the direction of the
relative momenta before and after the collision respectively and $\vec{n}%
\equiv \hat{k}\times \hat{q}$; $\vec{\epsilon}$ stands for the polarization
of the incident photon. The coefficients $F_{j}$ are linear combinations of
the partial-wave matrix elements multiplied by spherical harmonics and
weighted with geometrical factors. As such, they are functions of the energy
of the system and scattering angle $\cos (\theta )\equiv \hat{k}\cdot \hat{q}
$ only; $\theta $ is the scattering angle of the kaon relative to the
incident photon beam direction, $\hat{k}$. The explicit expressions for
these coefficients will be given elsewhere. It should be noted that the spin
structure given in Eq.(\ref{SSTRUC_P}) is equivalent to that of Ref.\cite%
{CGLN}.

Similarly, for a negative parity $\Theta^+$, we obtain 
\begin{equation}
\hat M^- = iG_1\ \vec\epsilon \cdot \hat q + G_2\ \vec\sigma \cdot
(\vec\epsilon \times \hat q) + G_3\ \vec\sigma \cdot (\vec\epsilon \times
\hat k) + G_4\ \vec\sigma \cdot \hat k \vec\epsilon \cdot \vec n + G_5\
[\vec\sigma \cdot \hat q \vec\epsilon \cdot \vec n + \vec\sigma \cdot \vec n
\vec\epsilon \cdot \hat q ]\ .  \label{SSTRUC_N}
\end{equation}
Quite recently, Zhao and Al-Khalili \cite{Zhao} have also given the spin
structure of the reaction amplitude for the case of negative parity $\Theta^+
$. The structure given above is equivalent to that of Eq.(18) in Ref.\cite%
{Zhao}, except for the term $\vec\sigma \cdot \vec n \vec\epsilon \cdot \hat
q$ in Eq.(\ref{SSTRUC_N}) which has not been included in Ref.\cite{Zhao} on
the grounds that it is a higher-order contribution. However, this term and
the $\vec\sigma \cdot \hat q \vec\epsilon \cdot \vec n$ term contribute with
the same coefficient $G_5(=iC_4)$ \cite{Zhao1}.

In what follows, $\vec{\epsilon}_{\perp }\equiv \hat{y}$ and $\vec{\epsilon}%
_{\parallel }\equiv \hat{x}$ denote the photon polarization perpendicular
and parallel to the reaction plane ($xz$-plane), respectively. Recall that
the reaction plane is defined as the plane containing the vectors $\vec{k}$
(in the $+z$-direction) and $\vec{q},$ and that $\vec{k}\times \vec{q}$ is
along the $+y$-direction. Then, from Eq.(\ref{SSTRUC_P}) 
\begin{eqnarray}
\hat{M}^{+\perp } &=&\alpha _{y}\ \sigma _{y}+i\alpha _{0}\sin (\theta )\ , 
\nonumber \\
\hat{M}^{+\parallel } &=&\alpha _{x}\ \sigma _{x}+\alpha _{z}\sin (\theta )\
\sigma _{z}\ ,  \label{PPOLM+}
\end{eqnarray}%
where 
\begin{equation}
\alpha _{0}\equiv F_{2}\ ,\ \ \ \alpha _{x}\equiv F_{1}+F_{4}\sin
^{2}(\theta )\ ,\ \ \ \alpha _{y}\equiv F_{1}\ ,\ \ \ \alpha _{z}\equiv
F_{3}+F_{4}\cos (\theta )\ .  \label{aux+}
\end{equation}%
Similarly, from Eq.(\ref{SSTRUC_N}) 
\begin{eqnarray}
\hat{M}^{-\perp } &=&\beta _{x}\ \sigma _{x}+\beta _{z}\sin (\theta )\sigma
_{z}\ ,  \nonumber \\
\hat{M}^{-\parallel } &=&\beta _{y}\ \sigma _{y}+i\beta _{0}\sin (\theta )\ ,
\label{PPOLM-}
\end{eqnarray}%
where 
\begin{eqnarray}
\beta _{0} &\equiv &G_{1}\ ,\ \ \ \ \ \ \ \ \ \ \ \ \ \ \ \ \ \ \ \ \ \ \ \
\ \ \ \ \ \ \ \ \ \ \ \ \ \ \beta _{x}\equiv G_{5}\sin ^{2}(\theta
)+G_{3}+G_{2}\cos (\theta )\ ,  \nonumber \\
\beta _{y} &\equiv &G_{5}\sin ^{2}(\theta )-G_{3}-G_{2}\cos (\theta )\ ,\ \
\ \ \ \ \beta _{z}\equiv G_{5}\cos (\theta )+G_{4}-G_{2}\ .  \label{aux-}
\end{eqnarray}

Eqs.(\ref{PPOLM+},\ref{PPOLM-}) exhibit an interesting feature in that the
Pauli spin structure of $\hat M^{+\perp}$ is the same as that of $\hat
M^{-\parallel}$, while the structure of $\hat M^{+\parallel}$ is the same as
that of $\hat M^{-\perp}$. This is a consequence of reflection symmetry.
Ultimately, we will exploit this feature to construct spin observables which
can determine the parity of $\Theta^+$ unambiguously.

We first consider the spin observables involving only the polarization of
the photon and/or nucleon in the initial state, for they are more easily
measured than observables involving the spin of $\Theta ^{+}$. For a given
photon polarization $\vec{\epsilon}_{\lambda }$, and target nucleon spin in
the $i$-direction ($i=x,y,z$), the corresponding spin-correlation
coefficient $A_{i}^{\lambda }$ can be expressed as 
\begin{equation}
\sigma ^{\lambda }A_{i}^{\lambda }=\frac{1}{2}Tr[\hat{M}^{\lambda }\sigma
_{i}\hat{M}^{\lambda \dagger }]\ ,  \label{BTxsc0}
\end{equation}%
where $\hat{M}^{\lambda }\equiv \sum_{m=0}^{3}M_{m}^{\lambda }\sigma _{m}$,
with $\sigma _{0}=1$, $\sigma _{1}=\sigma _{x}$, etc., denotes any of the $%
\hat{M}^{\lambda }$ (parity index $\pm $ suppressed) given in Eqs.(\ref%
{PPOLM+},\ref{PPOLM-}). The coefficients $M_{m}^{\lambda }$ can be read off
from these equations. $\sigma ^{\lambda }\equiv
\sum_{m=0}^{3}|M_{m}^{\lambda }|^{2}$ is the cross section with the
polarization of the photon $\vec{\epsilon}_{\lambda }$ incident on an
unpolarized target. Carrying out the trace in Eq.(\ref{BTxsc0}) yields 
\begin{equation}
\sigma ^{\lambda }A_{i}^{\lambda }=2\mathit{Re}[M_{0}^{\lambda
}M_{i}^{\lambda \ast }]+2\mathit{Im}[M_{j}^{\lambda }M_{k}^{\lambda \ast }]\
,  \label{BTxsc1}
\end{equation}%
where the subscripts $(i,j,k)$ run cyclically, i.e., (1,2,3), (2,3,1),
(3,1,2). In terms of individual cross sections $A_{i}^{\lambda }$ may be
written as 
\begin{equation}
A_{i}^{\lambda }=\frac{\sigma _{i}^{\lambda }(+)-\sigma _{i}^{\lambda }(-)}{%
\sigma _{i}^{\lambda }(+)+\sigma _{i}^{\lambda }(-)},  \label{BTxsc2}
\end{equation}%
where $\sigma _{i}^{\lambda }(+/-)$ denotes the cross section when photons
with polarization $\vec{\epsilon}_{\lambda }$ are incident on a target
nucleon with spin in the (positive/negative) $i$-direction.

Similarly, the target nucleon asymmetry, $A_i$, obtained using an
unpolarized photon beam on a target nucleon polarized in the $i$-direction
is given by 
\begin{eqnarray}
\sigma_u A_i & = & \frac{1}{2} Tr[\hat M \sigma_i \hat M^\dagger]  \nonumber
\\
& = & \sum_\lambda\left( 2\mathit{Re}[M^\lambda_0 M^{\lambda*}_i] + 2\mathit{%
Im}[M^\lambda_j M^{\lambda*}_k] \right) =\sum_\lambda \sigma^\lambda
A^\lambda_i \ ,  \label{Txsc}
\end{eqnarray}
where $\sigma_u \equiv \sum_\lambda \sigma^\lambda$ denotes the completely
unpolarized cross section; again, the subscripts $(i, j, k)$ run cyclically.
In the above equation, the first equality in the second row follows from
Eqs.(\ref{PPOLM+},\ref{PPOLM-}). In terms of individual cross sections $A_i$
may be written as 
\begin{equation}
A_i = \frac{\sigma_i(+) - \sigma_i(-)}{\sigma_i(+) + \sigma_i(-)}
\label{Txsc2}
\end{equation}
where $\sigma_i(+/-)$ denotes the cross section when unpolarized photons are
incident on a target nucleon with spin in the (positive/negative) $i$%
-direction.

We also consider the (linear) photon asymmetry given by 
\begin{equation}
\Sigma \equiv \frac{\sigma^\perp - \sigma^\parallel} {\sigma^\perp +
\sigma^\parallel} \ .  \label{PHASYM}
\end{equation}

Using Eqs.(\ref{PPOLM+},\ref{PPOLM-}) we find no \textit{model-independent}
way of relating the spin observables in Eqs.(\ref{BTxsc1},\ref{Txsc},\ref%
{PHASYM}), which are associated with only a polarized beam and/or target, to
the parity of the $\Theta^+$. Here, what could happen at best is that, by
constraining the kinematics of the reaction, one of these observables might
exhibit a markedly different angular dependence for the two choices of the
parity of $\Theta^+$. Note that when the coefficients $F_j$ and $G_j$ in
Eqs.(\ref{SSTRUC_P},\ref{SSTRUC_N}) are expanded in partial waves, their
angular dependences become explicit. It could also happen that, by
constraining the kinematics, one of these observables vanishes for one of
the choices of the parity of $\Theta^+$. If this is the case and the
corresponding measurement yields a non-vanishing value, we would know the
parity of $\Theta^+$. We have investigated these possibilities by
restricting the reaction to near-threshold kinematics and considering only $S
$- and $P$-waves in the final state. In this case, the coefficients $F_4$
and $G_5$ in Eqs.(\ref{SSTRUC_P},\ref{SSTRUC_N}) vanish, for they only
contain partial waves higher than the $P$-wave in the final state \footnote{$%
S$-waves contribute only to the coefficients $F_1$ and $G_3$ in Eqs.(\ref%
{SSTRUC_P},\ref{SSTRUC_N}).}. Unfortunately, none of these three spin
observables was found to exhibit the features described above.

The above considerations exhaust the spin observables involving only
polarization of the photon and/or nucleon in the \textit{initial} state and
show that these observables are unable to determine the parity of $\Theta^+$
in a model-independent way.

We now turn our attention to spin observables which also involve the
measurement of the polarization of $\Theta ^{+}$. These observables are
particularly suited to the use of Bohr's theorem \cite{Bohr}. This theorem
is a consequence of the invariance of the transition amplitude under
rotation and parity inversion and, in particular, reflection symmetry in the
scattering plane, and takes the form \cite{Satchler} 
\begin{equation}
\pi _{fi}=(-)^{M_{f}-M_{i}}\ .  \label{INV}
\end{equation}%
$\pi _{fi}$ denotes the product of the total intrinsic parity of the initial
and final states and $M_{(f/i)}$ denotes the sum of the spin projections in
the (final/initial) state along an axis normal to the scattering plane,
i.e., the $y$(or $\vec{k}\times \vec{q}$)-axis. Eq.(\ref{INV}) must be
satisfied by all parity-allowed transitions. For example, in the present
case, if the parity of $\Theta ^{+}$ is positive, then we must have $%
(-)^{M_{f}-M_{i}}=+1$, while if it is negative, $(-)^{M_{f}-M_{i}}=-1$.

We now exploit the reflection symmetry as manifested in Eq.(\ref{INV}) and
consider the (linear) photon asymmetry in conjunction with the polarization
transferred from the target nucleon to the $\Theta ^{+}$ which is given by: 
\begin{equation}
\Sigma _{yy}(i,j)\equiv \frac{\sigma _{y}^{\perp }(i,j)-\sigma
_{y}^{\parallel }(i,j)}{\sigma _{y}^{\perp }(i,j)+\sigma _{y}^{\parallel
}(i,j)}\ ,  \label{K_y}
\end{equation}%
where $\sigma _{y}^{(\perp /\parallel )}(i,j)$ stands for the cross section
for the photon polarization $\vec{\epsilon}_{(\perp /\parallel )}$ and the
spin orientation $i$ ($j$) of the nucleon ($\Theta ^{+}$) [up/down as $%
i,j=+/-$] along the $y$-axis. As mentioned above, it is easily verified from
Eq.(\ref{INV}) that, for the positive parity case, only spin-aligned
transitions ($i=j$) contribute to $\sigma _{y}^{\perp }(i,j)$ while only the
spin anti-aligned transitions ($i\neq j$) contribute to $\sigma
_{y}^{\parallel }(i,j)$. It follows from Eq.(\ref{INV}) that this feature is
just reversed in the case of a negative parity $\Theta ^{+}$. (Eqs.(\ref%
{PPOLM+},\ref{PPOLM-}) are consistent with these results as they should be.)
As a consequence, 
\begin{equation}
\Sigma _{yy}(j,j)=-\Sigma _{yy}(j,-j)=\pi _{\Theta }\ ,  \label{K^+_y}
\end{equation}%
where $\pi _{\Theta }$ stands for the parity of $\Theta ^{+}$. This result
is completely model independent and holds for any kinematic condition 
\footnote{%
These same model (and kinematic) independent considerations can be used to
relate the parity of the $\Theta ^{+}$ to spin observables in other
reactions. For example, in the $pp\rightarrow \Sigma ^{+}\Theta ^{+}$
reaction, $\pi _{\Theta }$ can be determined directly from 
\begin{equation}
\frac{\sigma _{y}(++,++)-\sigma _{y}(+-,++)}{\sigma _{y}(++,++)+\sigma
_{y}(+-,++)}=\pi _{\Theta }\ ,  \nonumber
\end{equation}%
where $\sigma _{y}(ij,kl)$ denotes the cross section with the spin
orientations $i$ and $j$ of the initial two protons [up/down as i,j=+/-]
along the $y$-axis and the spin orientations $k$ and $l$ along the $y$-axis
of the outgoing $\Sigma ^{+}$ and $\Theta ^{+}$, respectively.}. It should
be emphasized that the result in Eq.(\ref{K^+_y}) is based on the assumption
that the $\Theta ^{+}$ is a spin-1/2 particle. If the spin of $\Theta ^{+}$
is regarded as unknown, Eq.(\ref{K^+_y}) takes the more general form $\Sigma
_{yy}(M_{f}-M_{i}$ even$)=-\Sigma _{yy}(M_{f}-M_{i}$ odd$)=\pi _{\Theta }$.
Therefore it is clear that $\Sigma _{yy}(M_{f}-M_{i})$ measures directly the
parity of the $\Theta ^{+}$ for an arbitrary spin.

Another quantity which is related directly to the parity of $\Theta ^{+}$ is
the spin-transfer coefficient induced by a linearly polarized photon beam, $%
K_{ij}^{\lambda }$, which is given by 
\begin{eqnarray}
\sigma ^{\lambda }K_{ij}^{\lambda } &=&\frac{1}{2}Tr[\hat{M}^{\lambda
}\sigma _{i}\hat{M}^{\lambda \dagger }\sigma _{j}]\ ,  \nonumber \\
&=&(2|M_{0}^{\lambda }|^{2}-\sigma ^{\lambda })\delta _{ij}+2\mathit{Re}%
[M_{i}^{\lambda }M_{j}^{\lambda \ast }]+2\epsilon _{ijk}\mathit{Im}%
[M_{k}^{\lambda }M_{0}^{\lambda \ast }]\ ,  \label{STRANS0}
\end{eqnarray}%
where $\epsilon _{ijk}$ denotes the Levi-Civita antisymmetric tensor and $%
(i,j,k)$ may take any of the values $(1,2,3)$. The diagonal terms reduce to 
\begin{equation}
\sigma ^{\lambda }K_{jj}^{\lambda }=|M_{0}^{\lambda }|^{2}+|M_{j}^{\lambda
}|^{2}-\sum_{k\neq j}|M_{k}^{\lambda }|^{2}\ .  \label{K-ii0}
\end{equation}%
In terms of the individual cross sections $K_{jj}^{\lambda }$ may be written
as 
\begin{equation}
K_{jj}^{\lambda }=\frac{[\sigma _{j}^{\lambda }(+,+)+\sigma _{j}^{\lambda
}(-,-)]-[\sigma _{j}^{\lambda }(+,-)+\sigma _{j}^{\lambda }(-,+)]}{[\sigma
_{j}^{\lambda }(+,+)+\sigma _{j}^{\lambda }(-,-)]+[\sigma _{j}^{\lambda
}(+,-)+\sigma _{j}^{\lambda }(-,+)]}\ ,  \label{K-ii0xsc}
\end{equation}%
where, as before, $\sigma _{j}^{\lambda }(+,-)$, for example, corresponds to
the cross section induced by a photon beam with polarization $\vec{\epsilon}%
_{\lambda }$ on a target nucleon spin in the positive$(+)$ $j$-direction and
leading to the outgoing $\Theta ^{+}$ spin in the negative$(-)$ $j$%
-direction. Given the spin structure of the amplitude, Eq.(\ref{K-ii0xsc})
is often helpful in determining the characteristics of $K_{jj}^{\lambda }$.
Exploiting the structure of the amplitudes given in Eqs.(\ref{PPOLM+},\ref%
{PPOLM-}), it is straightforward to obtain 
\begin{equation}
K_{yy}^{\perp }=\pi _{\Theta }\ ,\ \ \ \ \ \ \ \ \ \ \ \ K_{yy}^{\parallel
}=-\pi _{\Theta }\ ,  \label{STRANSFER}
\end{equation}%
which are also model-independent results and hold for any kinematic
condition. It is also immediate that Eq.(\ref{K-ii0}) together with Eqs.(\ref%
{PPOLM+},\ref{PPOLM-}) yields $K_{xx}^{\parallel }=\pi _{\Theta }$ in
collinear kinematics or near-threshold \footnote{%
Note that in this kinematic condition, only the terms $\alpha _{(x/y)}$ and $%
\beta _{(x/y)}$ are non-vanishing in Eqs.(\ref{PPOLM+},\ref{PPOLM-}), for
all the coefficients in Eqs.(\ref{SSTRUC_P},\ref{SSTRUC_N}) vanish except $%
F_{1}$ and $G_{3}$.}. Apart from a minus sign, this result corresponds to
one of the results obtained recently in Ref.\cite{Rekalo}, i.e., [Eq.(8) in 
\cite{Rekalo}]. 

An alternative way to determine the parity of $\Theta ^{+}$ is in terms of
the spin-transfer coefficient using an unpolarized photon beam defined,
similar to Eq.(\ref{STRANS0}), by 
\begin{eqnarray}
\sigma _{u}K_{ij} &=&\frac{1}{2}Tr[\hat{M}\sigma _{i}\hat{M}^{\dagger
}\sigma _{j}]\ ,  \nonumber \\
&=&(2|M_{0}|^{2}-\sigma _{u})\delta _{ij}+2\mathit{Re}[M_{i}M_{j}^{\ast
}]+2\epsilon _{ijk}\mathit{Im}[M_{k}M_{0}^{\ast }]\ ,  \label{STRANS}
\end{eqnarray}%
where $(i,j,k)$ may take any of the values $(1,2,3)$ as in Eq.(\ref{STRANS0}%
). The diagonal terms reduce to 
\begin{equation}
\sigma _{u}K_{jj}=|M_{0}|^{2}+|M_{j}|^{2}-\sum_{k\neq j}|M_{k}|^{2}\ ,
\label{K-ii}
\end{equation}%
with $|M_{i}|^{2}=\sum_{\lambda }|M_{i}^{\lambda }|^{2}$. One can then,
immediately relate $K_{yy}$ to the (linear) photon asymmetry given by Eq.(%
\ref{PHASYM}), 
\begin{equation}
K_{yy}=\pi _{\Theta }\Sigma \ ,  \label{KPS}
\end{equation}%
which shows that by measuring both the spin transfer coefficient and photon
asymmetry, one can determine the parity of $\Theta ^{+}$ unambiguously. This
relation holds also for any kinematic condition and it was pointed out
recently by Rekalo and Tomasi-Gustafsson \cite{Rekalo} as a possible method
to pin down the parity of $\Theta ^{+}$.

Another way of determining the parity of $\Theta ^{+}$ is by measuring two
double-polarization observables, namely, the spin-correlation coefficient, $%
A_{i}^{\lambda }$, given by Eq.(\ref{BTxsc0}) and the polarization, $%
P_{i}^{\lambda }$, of the outgoing $\Theta ^{+}$ in the $i$-direction
induced by a photon beam with polarization $\vec{\epsilon}_{\lambda }$. The
latter is given by 
\begin{eqnarray}
\sigma ^{\lambda }P_{i}^{\lambda } &=&\frac{1}{2}Tr[\hat{M}^{\lambda }\hat{M}%
^{\lambda \dagger }\sigma _{i}]  \nonumber \\
&=&2\mathit{Re}[M_{0}^{\lambda }M_{i}^{\lambda \ast }]-2\mathit{Im}%
[M_{j}^{\lambda }M_{k}^{\lambda \ast }]\ ,  \label{BPxsc0}
\end{eqnarray}%
where the subscripts $(i,j,k)$ run cyclically. An analogous relationship to
that given by Eq.(\ref{BTxsc2}) also holds for $P_{i}^{\lambda }$, except
that the argument $(+/-)$ of $\sigma _{i}^{\lambda }$ now refers to the spin
orientation of the outgoing $\Theta ^{+}$ along the $i$-direction.
Exploiting the feature exhibited in Eqs.(\ref{PPOLM+},\ref{PPOLM-}), it is
straightforward to obtain 
\begin{equation}
A_{y}^{\perp }=\pi _{\Theta }P_{y}^{\perp }\ ,\ \ \ \ \ \ \ \ \ \ \ \
A_{y}^{\parallel }=-\pi _{\Theta }P_{y}^{\parallel }\ .  \label{C-P}
\end{equation}%
These results are again completely model independent and hold for any
kinematic condition.

Now, the two results in Eq.(\ref{C-P}) may be combined to yield 
\begin{equation}
\sigma^\perp A^\perp_y - \sigma^\parallel A^\parallel_y = \pi_\Theta
\sigma_u P_y \ ,  \label{CC-P}
\end{equation}
where $P_i$ denotes the polarization of the outgoing $\Theta^+$ in the $i$%
-direction induced by an unpolarized photon beam incident on an unpolarized
target nucleon; it is given by 
\begin{eqnarray}
\sigma_u P_i & = & \frac{1}{2} Tr[\hat M \hat M^\dagger \sigma_i]  \nonumber
\\
& = & \sum_\lambda\left( 2\mathit{Re}[M^\lambda_0 M^{\lambda*}_i] - 2\mathit{%
Im}[M^\lambda_j M^{\lambda*}_k] \right) =\sum_\lambda \sigma^\lambda
P^\lambda_i \ ,  \label{BPxsc1}
\end{eqnarray}
where, again, the subscripts $(i, j, k)$ run cyclically. The first equality
in the second row follows from Eqs.(\ref{PPOLM+},\ref{PPOLM-}). Note that in
Eq.(\ref{CC-P}), the r.h.s. of the equality involves the single polarization
observable, $P_y$, which may be easier to measure than the corresponding
double polarization observable, $P^\lambda_y$.

Yet, another possibility of determining the parity of $\Theta^+$ is to
measure two single polarization observables, namely, the target nucleon
asymmetry, $A_i$, given by Eq.(\ref{Txsc}) and the polarization of the
outgoing $\Theta^+$, $P_i$, given by Eq.(\ref{BPxsc1}). Using Eqs.(\ref%
{PPOLM+},\ref{PPOLM-}), we now form appropriate combinations of them, giving 
\begin{eqnarray}
A^+_y - P^+_y & = & 0 \ ,  \nonumber \\
A^-_y - P^-_y & = & 4\mathit{Im}[\beta_z\beta_x^*]\sin(\theta)/\sigma_u \ ,
\label{A-P}
\end{eqnarray}
for positive and negative parity $\Theta^+$, respectively. Again, the above
results are completely model independent and hold for any kinematic
conditions. However, unlike Eqs.(\ref{K^+_y},\ref{STRANSFER},\ref{KPS},\ref%
{C-P},\ref{CC-P}), the distinction between the positive and negative parity $%
\Theta^+$ is made by exclusion: if the measurement of $A_y - P_y$ yields a
non-vanishing value, the parity of $\Theta^+$ must be negative. Nothing can
be said about its parity, however, if the measurement yields a null value.

Obviously, measurements of any of the spin observables discussed in this
work (which can determine the parity of $\Theta ^{+}$) pose an enormous
experimental challenge, for they require measuring the spin of $\Theta ^{+}$
through its decay products $K+N$, in addition to the spin of the target
nucleon and/or photon. Furthermore, one also needs to consider the
background contribution which may potentially hinder the interpretation of
the required measurements, especially if the parity of $\Theta ^{+}$ happens
to be negative \cite{Nak}.

In summary, based on reflection symmetry in the scattering plane as encoded
either in Bohr's theorem [Eq.(\ref{INV})] or in the explicit forms of the
scattering amplitudes [Eqs.(\ref{PPOLM+},\ref{PPOLM-})], we have
demonstrated that some spin observables in $\Theta ^{+}$ photoproduction can
be related directly to the parity of $\Theta ^{+}$. In particular, Eqs.(\ref%
{K^+_y},\ref{STRANSFER},\ref{KPS},\ref{C-P},\ref{CC-P}) offer ways of
providing a model-independent determination of the parity of $\Theta ^{+}$.
Also, we have shown that measurements of the target nucleon asymmetry and
the $\Theta ^{+}$ polarization induced using an unpolarized photon beam [Eq.(%
\ref{A-P})] may be useful in determining the parity of $\Theta ^{+}$ in a
model-independent way. Furthermore, we have also shown that, in this
reaction, no spin observables involving only the polarization of the photon
and/or nucleon in the initial state can determine the parity of $\Theta ^{+}$
unambiguously. Finally, because of its generality, Bohr's theorem [Eq.(\ref%
{INV})] may, of course, be used in a similar way to analyze other reactions
induced by photons or other probes.

\vspace{2em} \noindent \textbf{Acknowledgements:}\newline
\noindent We thank Qiang Zhao for pointing out the equivalence of the
negative parity amplitude used here with that given in Ref.\cite{Zhao}. This
work is supported by Forschungszentrum-J\"{u}lich, contract No. 41445282
(COSY-058).


%

\end{document}